\documentstyle [a4,12pt]{article}
\newcommand{\mincir}{\raise -2.truept\hbox{\rlap{\hbox{$\sim$}}\raise5.truept
\hbox{$<$}\ }}
\newcommand{\magcir}{\raise -2.truept\hbox{\rlap{\hbox{$\sim$}}\raise5.truept
\hbox{$>$}\ }}
\newcommand{\minmag}{\raise-2.truept\hbox{\rlap{\hbox{$<$}}\raise 6.truept\hbox
{$>$}\ }}
\newcommand{\be}{\begin{equation}}
\newcommand{\ee}{\end{equation}}
\newcommand{\ba}{\begin{eqnarray}}
\newcommand{\ea}{\end{eqnarray}}
\newcommand{\brr}{\begin{array}}
\newcommand{\err}{\end{array}}
\newcommand{\bc}{\begin{center}}
\newcommand{\ec}{\end{center}}

\newcommand{\et}{et al.~}

\newcommand{\gsim}{\ \rlap{\raise 2pt \hbox{$>$}}{\lower 2pt \hbox{$\sim$}}\ }
\newcommand{\lsim}{\ \rlap{\raise 2pt \hbox{$<$}}{\lower 2pt \hbox{$\sim$}}\ }
\newcommand{\vm}{ { \{{\vec v}^{(m)}           \} }  }

\title{ {\bf OBSERVATIONAL CONSTRAINTS ON BLUE PRIMORDIAL SPECTRA}}
\author{
{\bf Francesco Lucchin}$^1$, {\bf Sergio Colafrancesco}$^2$, \\
{\bf Giancarlo de Gasperis}$^3$, {\bf Sabino Matarrese}$^4$,
{\bf Simona Mei}$^5$, \\
{\bf Silvia Mollerach}$^6$, {\bf Lauro Moscardini}$^1$,
{\bf Nicola Vittorio}$^7$ \\ ~\\
{\em $^1$Dipartimento di Astronomia, Universit\`a di Padova,}\\
{\em vicolo dell'Osservatorio 5, I--35122 Padova, Italy} \\ \\
{\em $^2$Osservatorio Astronomico di Roma,} \\
{\em via dell'Osservatorio 5, I--00040 Monteporzio (RM), Italy}\\ \\
{\em $^3$Dipartimento di Fisica, Universit\`a dell'Aquila,} \\
{\em via Vetoio 1, I--67010 Coppito (AQ), Italy} \\ \\
{\em $^4$Dipartimento di Fisica G. Galilei, Universit\`a di Padova,} \\
{\em via Marzolo 8, I--35131 Padova, Italy} \\ \\
{\em $^5$Universit\'e Louis Pasteur, Observatoire Astronomique de
Strasbourg,} \\
{\em 11 rue de l'Universit\'e, F--67000 Strasbourg, France} \\ \\
{\em $^6$SISSA -- International School for Advanced Studies,}\\
{\em via Beirut 2--4, I--34013 Trieste, Italy} \\ ~\\
{\em $^7$Dipartimento di Fisica, Universit\`a di Roma ``Tor Vergata",}\\
{\em via della Ricerca Scientifica, I--00133 Roma , Italy}\\ \\}

\date{}

\begin{document}

\maketitle

\newpage
\section*{Abstract}
Recently, there has been growing interest on primordial ``blue" ($n>1$)
perturbation spectra, motivated both by a composite set of observational data
on large scales and from the point of view of theoretical model building. After
reviewing the theoretical (inflationary) motivations for these blue spectra, we
consider various observational constraints, within both Cold Dark Matter and
Mixed Dark Matter scenarios. In particular, using linear theory, we discuss
large--scale bulk flows and the X--ray cluster abundance. We also perform
various N--body simulations of these models to study the clustering properties
of the matter distribution and the peculiar velocity field.

\section{Introduction}

The anisotropy of the Cosmic Microwave Background (CMB), detected on large
angular scales by {\it COBE} (Smoot et al. 1992; Bennett et al. 1994), provides
an insight on primordial perturbations. This measurement yields, on one hand,
the amplitude of density perturbations at the largest observable scales, and,
on the other hand, the spectral index $n$, that fully determines the
power--spectrum [i.e. $P(k) \propto A k^n$] of the primordial density
fluctuation field, up to a possible contribution by a stochastic background of
gravitational waves. These constraints, promptly taken into account in the
current cosmological debate, can discriminate among the possible scenarios. In
particular, the standard (i.e. $n = 1$) Cold Dark Matter (hereafter SCDM) model
with the {\it COBE} normalization predicts too much structure on scales $\lsim
10 ~h^{-1}$ Mpc ($h$ being the Hubble constant in units of 100 km s$^{-1}$
Mpc$^{-1}$), e.g. in terms of an excessive abundance of clusters (e.g. White,
Efstathiou, \& Frenk 1993; Colafrancesco \& Vittorio 1994), and of too a large
rms pairwise galaxy velocities on Mpc scales (Davis et al. 1985; see however
Mo, Jing, \& Borner 1993; Zurek et al. 1994). Tilted (i.e. $n<1$) CDM models
are in better agreement with observations on small scales, but the amount of
tilting is strongly constrained from data on the large--scale velocity field
(Moscardini et al. 1995) and by cluster clustering properties (Plionis et al.
1995). The most severe flaw of tilted CDM models is the very late epoch of
galaxy formation.

A fashionable alternative to SCDM, recently emerged to overcome these problems,
is provided by Mixed (i.e. cold + hot) Dark Matter (MDM) models: adding to the
cold dark matter a small contribution ($\sim 10-30 \%$ of the critical density)
of massive neutrinos erases, via free--streaming, part of the power on the
scales of galaxies and clusters. This scenario will also be reinforced from the
particle physics side, if recent claims for a mass of order $2$ eV for the
$\mu$ neutrino will be confirmed (Louis 1994). The standard MDM model is based
on the same assumptions as SCDM: in particular, the primordial power--spectrum
is assumed again to be scale--invariant (i.e. $n=1$). Although the results of
linear analyses and numerical simulations (e.g. Klypin et al. 1993; see also
Davis, Summers, \& Schlegel 1992; Jing et al. 1994; Cen \& Ostriker 1994;
Klypin, Nolthenius, \& Primack 1995) are certainly promising, some problems are
clearly present: an example is the excessive delay of the epoch of galaxy
formation, in contrast with observations of quasars and damped Ly$\alpha$
system at redshifts $z \geq 3$ (Ma \& Bertschinger 1994; Klypin et al. 1995).

The analysis of the first two years of the {\it COBE} DMR data indicates a most
likely value for the spectral index in the range $n=0.9 - 1.9$ (Bennett et al.
1994; see also G\'orski et al. 1994), a value also suggested by the Tenerife
experiment at $5^\circ$ (Hancock et al. 1994) and by the MAX measurements at
$1^\circ$ (e.g. Gundersen et al. 1993; Devlin et al. 1994). An $n>1$
perturbation spectrum can also be at the origin of the claimed bulk flow on
scales of 150 $h^{-1}$ Mpc (Lauer \& Postman 1994) and of the large voids in
the galaxy distribution of the CfA survey (de Lapparent, Geller, \& Huchra
1986; Geller \& Huchra 1989), at scales of order 50 $h^{-1}$ Mpc, as pointed
out by Piran et al. (1993). On the other hand, it is also possible to put upper
bounds on the value of $n$. Carr \& Lidsey (1993) found $n<1.5$, from
primordial black hole overproduction constraints on nucleosynthesis and the
{\it COBE} data (see also Carr, Gilbert \& Lidsey 1994). Hu, Scott, \& Silk
(1994) set an independent constraint, $n<1.54$, by using recent experimental
limits on the CMB spectral distortions found by {\it COBE} FIRAS.

Most theories for the origin of the primordial fluctuations predict a spectral
index $n\leq 1$. In particular, inflationary models based on simple inflaton
potentials originate this type of spectra. Nevertheless, in recent years some
authors have pointed out that suitable inflationary dynamics can easily account
for blue spectra (Liddle \& Lyth 1993; Linde 1994; Mollerach, Matarrese, \&
Lucchin 1994; Copeland et al. 1994), in particular, in the framework of the
so--called hybrid models.

In this paper we investigate the predictions of a possible variant of MDM
models, by assuming a ``blue", i.e. $n>1$, primordial density perturbation
spectrum; this BMDM model (where ``B" now stands for ``blue"), suitably
normalized to {\it COBE} DMR, may alleviate the small--scale problem of the
standard (i.e. $n=1$) MDM model. An important advantage of ``anti--tilting" the
spectrum is that it provides an earlier galaxy formation epoch. Here we
restrict ourselves to discuss the present matter and light distribution, both
in the linear and nonlinear regime. We refer to de Bernardis, Baldi, \&
Vittorio (1995) for a discussion of how MDM models compare with the anisotropy
data at degree angular scales.

The plan of the paper is as follows. In Section 2 we discuss the theoretical
motivation for blue power--spectra from the inflationary point of view. In
Section 3 we show the predictions of the linear theory for the large--scale
peculiar velocities and for the galaxy cluster abundance. In Section 4 we
present the analysis of our N--body simulations. Finally, in Section 5 we
discuss our results and draw our main conclusions.

\section{Inflationary models}

Quantum fluctuations of scalar fields during inflation provide a very appealing
explanation for the origin of the primordial density perturbations. It is
useful to describe these perturbations in terms of the gauge--invariant
variable $\zeta$. Its power--spectrum  to first order in the slow roll
approximation is given by
\be
P_{\zeta} ^{1/2} (k) = \left. { H^2  \over 2 \pi \dot \phi}
\right|_{aH = k}.
\ee
Its spectral index, defined by $\alpha \equiv d \ln P_\zeta^{1/2}/ d \ln k$, is
related to the spectral index of density perturbations on constant time
hypersurfaces before matter--radiation equality as $n = 2 \alpha + 1$. Most
inflationary models predict a spectral index $n < 1$. This is the case for
example for models based on a polynomial potential, which  give rise to a
power--spectrum $P_\zeta ^{1/2} (k)$ logarithmically decreasing with $k$.
Another example is the power--law inflation, which is based on an exponential
potential and gives rise to a power--spectrum decreasing with $k$ as a power
(Lucchin \& Matarrese 1985). This is due to the fact that in most inflationary
models $H^2/\vert\dot \phi\vert$ decreases with time as the inflaton evolves,
thus the larger wavenumber perturbations, that left the horizon later, have
smaller amplitude. However, as it has recently been discussed by Mollerach,
Matarrese, \& Lucchin (1994), there is a class of simple inflationary models
leading to density perturbations with $n > 1$ (see also Liddle \& Lyth 1993;
Carr \& Lidsey 1993; Linde 1994; Copeland et al. 1994). These are produced
while the inflaton field rolls down to a potential minimum with potential
energy dominated by a cosmological constant--like piece. In this case
$\vert\dot \phi \vert$ decreases as $\phi$ approaches the potential minimum,
while $H^2$ remains approximately constant in such a way that $H^2/\vert\dot
\phi\vert$ increases with time. The shape of the inflaton potential leading to
$P_\zeta^{1/2} (k) \propto k^\alpha$ with $\alpha =$ const $ >0$ has been
obtained by Mollerach, Matarrese, \& Lucchin (1994); it looks quite
complicated. However, if the condition that $\alpha$ is constant for all scales
is relaxed, a huge class of potentials with a non--vanishing energy density in
the minimum, which corresponds to an effective cosmological constant, can work.
For example, the potentials $V(\phi) = V_0 + f \phi^m$ give rise to a spectrum
decreasing with $k$ for wavelengths that left the Hubble radius when $f \phi^m
> V_0 $,
\be
n \simeq 1 + {2 f m (m-1)\over \kappa^2 V_0} \phi^{m-2},
\label{n}
\ee
where the right hand side has to be evaluated at the time when the wavelengths
of interest left the horizon during inflation. In particular, we see that the
quadratic potential case, $m = 2$, gives rise to an approximately constant
spectral index $n > 1$.

One can think that these kinds of potentials will have problems to end
inflation and reheat the universe, as the two usual reheating mechanisms,
namely rapid oscillations of the inflaton field around the potential minimum
and first--order phase transitions with bubble production do not work in this
case. This reheating problem can nevertheless be avoided by the ``hybrid
inflation'' mechanism, recently proposed by Linde (1991a,b, 1994). According to
it, inflation ends in the following way: the slow roll of the inflaton field at
a given time triggers a second--order phase transition of a second field, whose
false vacuum energy density is responsible for the cosmological constant term.
The field rolls down to its potential minimum and oscillates around it. A
particular potential in which this scenario can be realized is (Linde 1991a,b,
1994)
\be
V(\phi,\sigma) = {1 \over 4\lambda} (M^2-\lambda\sigma^2)^2 +{m^2
\over 2}\phi^2 +{g^2 \over 2}\phi^2\sigma^2.
\ee
For $ \phi > \phi_c \equiv M/g$, $\sigma$ is in its false vacuum at $\sigma =
0$. The potential for $\phi$ in this regime is given by $ V(\phi) = M^4
/4\lambda + m^2 \phi^2 /2$. As we have discussed, when the cosmological
constant term dominates, this gives rise to an inflationary phase with a
power--spectrum $P_\zeta (k)$ increasing with $k$. From eq.(\ref{n}), the
spectral index for the perturbations is given by $\alpha \simeq 4 \lambda m^2 /
\kappa^2 M^4$. To be consistent with the approximations done, this result holds
only for small values of $\alpha$. For a given value of $\alpha$, this fixes
the relation between $m^2$ and $M^4$. When $\phi < \phi_c$, $\sigma$ rolls down
to its minimum and the universe reheats.

There are some  constraints on the model parameters in order that it works. In
order to have at least 60 e--foldings of inflation for $\phi > \phi_c$, one
needs $M^3 \gsim m^2\lambda e^{60\alpha} M_P/g$, and in order that inflation
finishes soon after $\phi$ reaches $\phi_c$, one needs $M^6\lsim 3 \lambda^2
M_P^4 m^2/ 2 \pi^2$.

The amplitude of the perturbations that left the horizon around 60 e--foldings
before the end of inflation (the ones relevant for structure formation) is
given by
\be
P_\zeta^{1/2} (k_{60})= {\sqrt{2\pi }g M^5 \over
\sqrt{3}\lambda^{3/2} M_P^3 m^2} \exp\left(-60 {4 \lambda m^2 \over
\kappa^2 M^4}\right).
\ee
We normalize its amplitude using the {\it COBE}  $Q_{{\rm rms-PS}}$
determination, $P_\zeta(k_{60}) \equiv A^2 (n) = (Q_{{\rm rms-PS}}/ T_0)^2 40
\Gamma^2(2-n/2) \Gamma(9/2-n/2)/\pi \Gamma(3-n)\Gamma(3/2+n/2)$. From the
analysis of the two first years of {\it COBE} data, $Q_{{\rm rms-PS}}= 18.2
e^{0.58(1-n)} \mu$K (Bennett et al. 1994) and the FIRAS value $T_0= 2.726 \pm
0.010$ K (Mather et al. 1994), the resulting values of $A(n)$ are of the order
of $4 \times 10^{-5}$ for the values of $n$ of interest. This normalization,
combined with the previous relation between $m^2$ and $M^4$ in terms of
$\alpha$, gives the values of $m$ and $M$:
\be
M\simeq 4.3 A(n) \alpha e^{60\alpha}{\sqrt{\lambda}\over g}
M_P,\ \ \ \ \ \ \
m \simeq 47 A^2(n) \alpha^{5/2} e^{120\alpha}
{\sqrt{\lambda}\over g^2} M_P.
\ee
For a given value of $\alpha$, these relations fix the values of $m$ and $M$.
We have to check that the constraints imposed on the model parameters are
satisfied. The condition $M^2\phi^2/2 <M^4/4\lambda$ translates to
$\lambda^{1/4}< 50 e^{-60\alpha} g/ \alpha^{3/4}$. We can ensure that this is
satisfied for all the values of $\alpha$ of interest taking for the coupling
constants $g \sim {\cal O}(1)$ and $\lambda$ small enough. The condition for
inflation to end soon after $\phi \sim \phi_c$ holds only for $\alpha \lsim
0.17$ ($n \lsim 1.35$). The additional constraint that the energy density be
smaller that the Planck density, $M^4/4\lambda < M_P^4$, always holds when the
above conditions are satisfied.

\section{Linear Tests of Blue Primordial Spectra}

Following the inflationary paradigm, we restrict ourselves to flat cosmological
models, where a Hubble parameter $h=0.5$ is assumed and baryons contribute only
for the $5 \%$ to the critical density (i.e. $\Omega_B=0.05$). The fraction of
the critical density in cold and hot dark matter is
$\Omega_{CDM}=1-\Omega_B-\Omega_{\nu}$ and $\Omega_\nu$, respectively. The
family of flat MDM model is then characterized only by the parameter
$\Omega_\nu$ that in this Section we let vary between $0$ (pure CDM) and $\lsim
0.5$ (the MDM model where hierarchical clustering still occurs).

In order to study the present matter distribution in MDM models, we evaluate
the transfer function of density fluctuations in each of the four components
(CDM, HDM, baryons, photons), by numerically integrating the full set of
equations for the perturbed matter--energy density field in the synchronous
gauge formalism. The details of our calculation are described in de Gasperis,
Muciaccia, \& Vittorio (1995; see also Peebles 1981; Bonometto et al. 1984).
The results for $\Omega_{\nu}=0.3$ are shown in Figure 1, where we plot the
transfer function $T(k)$ for each separate component: CDM, HDM and baryons,
respectively. Here we assume that there is only one family of massive
neutrinos. So, $\Omega_\nu=0.3$ corresponds to a neutrino mass $m_{\nu} \approx
7 {\rm eV} $. We also plot the total transfer function defined as follows:
\be
T_{TOT}(k) = \Omega_{CDM} T_{CDM}(k) + \Omega_{\nu} T_{\nu}(k) +
\Omega_B T_{B}(k) \ .
\ee
The transfer function is given in arbitrary units. However the relative
amplitude of $T(k)$ at different redshifts reflects the linear growth of
density fluctuation in an Einstein--de Sitter universe.  We evaluated the
transfer functions at redshift $z=20$, in order to properly start our N--body
calculations (see next section), with the right statistical segregation among
the different components. In Figure 1 it is evident that baryons (dotted line)
catch up completely the CDM (dashed line) at the present time. The transfer
function of massive neutrinos is initially lower than the CDM one because of
free--streaming. However, it gets presently comparable to that of the other
components, at least for the range of wavenumbers shown in Figure 1.

Figure 2 shows the rms value of the density fluctuation field at the present
time, $\sigma(R)$, smoothed with a top--hat filter of radius $R$. We plot
$\sigma(R)$ for the following models: SCDM, a blue (with $n=1.2$) CDM model
(hereafter BCDM), MDM and BMDM ($n=1.2$) with $\Omega_{\nu}=0.3$. All models
are normalized to the rms value ($30~\mu$K) of the temperature fluctuations in
the DMR maps. It is apparent the reduction of power on small scales and the
flattening of $\sigma(R)$ for the MDM model compared to SCDM. Our BMDM model is
intermediate between SCDM and MDM. So, this model can in principle be
consistent with the small--scale clustering of galaxies. This alleviates the
problem of the standard MDM model, where structure formation is expected to
occur later than in SCDM. To roughly quantify this issue we evaluate the
redshift of formation of structures by asking when, on a given scale, a one
sigma fluctuation goes nonlinear. In Table 1 we show the redshifts of
nonlinearity for three mass scales ($M= 10^{12},~ 10^{13}$, and $10^{14} ~{\rm
M}_{\odot}$), corresponding to typical masses of galaxies ($z_{nl,gal}$),
groups ($z_{nl,gr}$) and clusters of galaxies ($z_{nl,cl}$), respectively.

\begin{center}
{\bf Table 1.} Redshift of Galaxy, Group and Cluster Formation
\end{center}
\begin{center}
\begin{tabular}{|l|r|r|r|}
\hline
\hline
        & $z_{nl,gal}$         & $z_{nl,gr}$   & $z_{nl,cl}$ \\
\hline
\hline
SCDM    & $4.4$  & $2.5$ & $1.0$ \\
BCDM    & $11.4$ & $6.5$ & $3.1$ \\
MDM     & $0.6$  & $0.3$ & $0.1$ \\
BMDM    & $1.9$  & $1.3$ & $0.7$ \\
\hline
\end{tabular}
\end{center}

\subsection{Bulk Velocities}

We test our models against peculiar velocity data. We consider the Local Group
(LG) velocity relative to the CMB (Kogut et al. 1993), assigning a velocity of
$|{\vec v}^{(1)}| = 627 \pm 22$ km s$^{-1}$ to a sphere of radius
$R_1=5~h^{-1}$ Mpc. We also use recent data by Dekel (1994), which give the
bulk flow $\vm$ ($m=2,7$) of spheres of radius $R_m=10,~20,~\dots ,~60~h^{-1}$
Mpc. These data were derived after reconstructing with the POTENT method
(Bertschinger et al. 1990) the three--dimensional velocity field smoothed by a
Gaussian window with a radius of $12 ~h^{-1}$ Mpc. Therefore, in the following
we will consider the set $\vm$ ($m=1,7$) corresponding to the LG velocity and
six bulk flows.

The likelihood of the data $p(\vm|{\cal H})$ under the hypothesis ${\cal H}$,
in our case a model with a given $\Omega_{\nu}$ and $n$, can be written as
\be
p(\vm|{\cal H}) =
{1 \over (2\pi)^{3N/2}} {1\over\vert V\vert ^{3/2}}
\exp\biggl[-{1\over 2}\sum_{i=1}^N \sum_{j=1}^N
V_{ij}^{-1} \bigl( \vec{v}^{(i)} \cdot \vec{v}^{(j)}\bigr)\biggl]\ ,
\ee
where $V_{ij}^{-1}$ and $\vert V\vert$ are the inverse and the determinant of
the $7\times 7$ correlation matrix, $V_{ij} \equiv \langle {\vec v}^{(i)} \cdot
{\vec v}^{(j)} \rangle/3$, respectively. The condition ${\cal H}$ is contained
in the elements of the correlation matrix, which can be written as
\be
V_{ij} ={H_0^2\over 6\pi^2} \int dk\,P(k){\widetilde W}_i(k R_i) {\widetilde
W}_j(k R_j) + {\Sigma_i^2 \over 3} \delta_{ij}\ ,
\ee
where the indices $i$ and $j$ refer to the velocities of spheres of different
radii as indicated in Table 2 and $P(k)$ is the density fluctuation
power--spectrum of the chosen theoretical model (properly normalized to {\it
COBE} DMR data). Moreover, in eq.(8), ${\widetilde
W}_i(kR_i)=3j_{1}(kR_i)/(kR_i)^3$.

The observational data $|{\vec v_{obs}}^{(i)}|$ with their error bars
$\Sigma_{obs}^{(i)}$ are compared in Table 2 with the theoretical predictions
for the three--dimensional rms bulk flows $|{\vec v}^{(i)}_{th}|$ [given by
$\sqrt{3 V_{ii}}$ ($i=1,7$)] obtained for  SCDM, BCDM (with $n=1.2$), MDM and
BMDM ($n=1.2$) with $\Omega_{\nu}=0.3$. The corresponding values of the bias
parameter $b \equiv \sigma^{-1} (8 h^{-1} {\rm Mpc})$ are also given in Table
2.

\newpage
\begin{center}
{\bf Table 2.} Bulk velocities. Observational data and theoretical predictions.
\end{center}
\begin{center}
\begin{tabular}{|l|r|r|r|r|r|r|r|}
\hline
\hline
& & & & SCDM & BCDM & MDM & BMDM \\
\hline
\hline
& & &  bias & 0.95 & 0.50 &1.47 & 0.91 \\
\hline
\hline
i & $R_i$ & $|{\vec v_{obs}}^{(i)}|$ & $\Sigma_{obs}^{(i)}$ & $|{\vec
v}^{(i)}_{th}|$ &
$|{\vec v}^{(i)}_{th}|$ & $|{\vec v}^{(i)}_{th}|$ & $|{\vec v}^{(i)}_{th}|$ \\
\hline
\hline
1& 5  & 627 &  22 &  880 & 1387 & 723 & 1107 \\
2& 10 & 494 & 170 &  509 &  705 & 488 &  670 \\
3& 20 & 475 & 160 &  467 &  638 & 452 &  614 \\
4& 30 & 413 & 150 &  418 &  562 & 409 &  548 \\
5& 40 & 369 & 150 &  377 &  496 & 371 &  488 \\
6& 50 & 325 & 140 &  338 &  438 & 335 &  434 \\
7& 60 & 300 & 140 &  310 &  394 & 308 &  391 \\
\hline
\end{tabular}
\end{center}

\bigskip
Now we consider models with $1 \le n \le 1.5$ and $\Omega_\nu \le 0.5$ and we
look for the model most consistent with the data, i.e. the one that maximizes
the Likelihood of the set $\vm _{obs}$. We find that this happens for
$\Omega_{\nu}=0$ and $n=1.1$, if we maximize the joint probability of having
the measured bulk flows {\it and} the CMB dipole. We find instead
$\Omega_{\nu}=0$ and $n=1.2$, if we maximize the conditional probability of
having the measured bulk flows, under the condition that the CMB dipole has its
measured value. The conditional probability is obtained by dividing eq.(7) by
\be
p({\vec v}^{(1)}|{\cal H}) = {1 \over (2\pi)^{3/2}} {1\over V_{11}^{3/2}}
\exp\biggl[-{1\over 2} {|\vec{v}^{(1)}|^2  \over V_{11} } \biggr]\ .
\ee
Thus, BCDM models seem to be favoured from this kind of analysis.

In order to define a $95\%$ confidence level for $\Omega_{\nu}$ and $n$ we
proceed as follows. We define a likelihood ratio $\lambda \equiv p(\vm|{\cal
H}_1)/p(\vm|{\cal H}_2)$, and we test the condition ${\cal H}_1$ (a model with
given $\Omega_{\nu}$ and $n$) against the condition ${\cal H}_2$ (the most
probable model). Here $p$ is either the joint [of Eq.(7)] or the conditional
probability of reproducing the observed set $\vm_{obs}$. We generate 10,000
realizations of the set $\vm$ under ${\cal H}_1$ and ${\cal H}_2$,
respectively, and from them we evaluate $p(\lambda|{\cal H}_1)$ and
$p(\lambda|{\cal H}_2)$ (see Del Grande \& Vittorio 1992 for more details on
this method). We find that BMDM models, with a small fraction of hot dark
matter, $\Omega_\nu \lsim 0.3$, can provide a reasonable agreement with the
peculiar velocity data. We will compare this conclusion with the analysis of
detailed N--body simulations in Section 4.2.

\subsection{Cluster Abundance: the X--Ray Constraints}

In this section we test the predictions of models with primordial blue
perturbation spectra against the local X--Ray Luminosity Function (XRLF) of
galaxy clusters (Kowalski et al. 1984). In hierarchical scenarios the XRLF can
be derived from the cluster mass distribution using the appropriate $M/L$
relation between the cluster mass $M$ and its X--ray luminosity $L$. Several
evidences (see e.g. Colafrancesco \& Vittorio 1994 for a discussion) suggest
that the cluster mass distribution can be reasonably represented by the Press
\& Schechter (1974) formula
\be
N(M,t)= {{\cal I} \over M} ~\bigg\vert {d\nu_c \over dM} \bigg\vert
{}~{\rho \over (2 \pi)^{1/2} } ~{\rm e}^{- \nu_c^2/2}~,
\ee
where $\nu_c \equiv \delta_c / \sigma(M,z)$, $\delta_c = 1.7$, and $\rho$ is
the mean density at time $t$ (the mass $M$ is defined through the corresponding
radius of a top--hat filter of radius $R$). The normalization is uncertain
within a factor $1\lsim {\cal I} \lsim 2$ (see e.g. Bond et al. 1991). A direct
comparison of the mass distribution in eq.(10) with the data of Bahcall \& Cen
(1992) is not straightforward, as the mass $M$ in the Press \& Schechter theory
can be related to the estimated mass within a fixed radius, $M_A= M(r< 1.5
{}~h^{-1}$ Mpc), only if detailed information about the cluster density
profiles
is available (e.g. Antonuccio--Delogu \& Colafrancesco 1994, for a discussion).
On the contrary, the X--ray luminosities are mostly sensitive to the inner
parts of the density profiles, as the specific emissivities [mainly due to
thermal bremmstrahlung from the hot intergalactic medium (IGM)] are
$\epsilon_{\nu} \sim y^{-\alpha} \exp(-y)$, where $y=h\nu/kT$ and $\alpha
\approx 0.3-0.4$ (Gioia et al. 1990). So, apart from the theoretical
difficulties in explaining the existence of a cluster core, the X--ray
luminosity $L_x \propto \int dr r^2 {\overline n}^2(r) T^{1/2}(r)$ is a more
reliable quantity than $M_A$, due to its relation to the total cluster mass
$M$. Here $\overline n$ is the density of the IGM at temperature $T$ and the
integral is extended over the cluster volume. We adopt here the following
parameterization for the local cluster X--ray luminosity
\be
L \propto  M^c \rho^d {\cal K}(\Delta E,z)\ ,
\ee
where $c=4/3$, $d=7/6$, and $ {\cal K} (\Delta E, z) = \int_{y}^{y+\Delta y} dy
{}~\epsilon_{\nu}(y) / \int_0^{\infty} dy ~\epsilon_{\nu}(y) $ is the
K--correction in the energy band $\Delta y$. This  correction, applied to the
HEAO1 energy band $2-6$ keV, yields, at $z \approx 0$, a reduction of the total
luminosities by a factor $ \sim 3$ (see Colafrancesco \& Vittorio 1994, for a
more complete discussion).

Given $N(M,z)$ and the ratio $M/L \propto M^{1-c}$ from eq.(11), we obtain the
comoving XRLF
\be
N(L,z) \equiv N(M,z) {dM \over dL} = {1 \over c} {M \over L} N\big[
M(L,z,\Delta E), z\big]\ .
\ee
We plot in Figure 3 the local XRLF predicted by our models, normalized to {\it
COBE} DMR, choosing the canonical value $\delta_c \approx1.7$, to select
collapsed objects. The XRLFs predicted in all our models are well above the
observational data, indicating that {\it COBE} normalized models are unable to
reproduce the cluster abundance, quite independently of the value of the
spectral index $n$. If we allow for $\delta_c$ values larger than $1.7$, we can
reproduce the observed XRLF at a better level (see Colafrancesco \& Vittorio
1994). In fact, the mass distribution in eq.(12) depends upon the combination
$\delta_cb$ and ${\cal I}$. So for a given model (i.e. for a given
$\Omega_{\nu}$ and $n$) we may derive $\delta_cb$ and ${\cal I}$ by a direct
fit to the XRLF data. The results of a Chi--square analysis are shown in Table
3. The theoretical XRLFs, obtained with the values of Table 3 for $\delta_cb$
and ${\cal I}$ are also shown in Figure 3. Note that for $\Omega_{\nu}=0.3$ and
$n=1$ the difference between the two approaches is quite small.

A discussion of the cluster abundance obtained in N--body simulations will be
presented in Section 4.3.

\begin{center}
{\bf Table 3.} Parameters of the XRLF fit
\end{center}
\begin{center}
\begin{tabular}{|c|l|l|l|}
\hline
\hline
($\Omega_{\nu},n$)  & $\delta_c b$  & ${\cal I}$  & $\chi^2_{min}$ \\
\hline
\hline
($0.1,1.3$) & $2.7$  & $1.3$  & $1.93$ \\
($0.2,1.4$) & $2.7$  & $1.3$  & $1.74$ \\
($0.3,1.4$) & $2.8$  & $1.7$  & $1.59$ \\
\hline
\end{tabular}
\end{center}

\section{N--body simulations}

To study the large--scale structure of the Universe in models with blue
primordial perturbation spectra we ran N--body simulations of the matter
distribution. We used a particle--mesh code with a box of side $260~h^{-1}$ Mpc
(we adopt here $h=0.5$). We considered three different models: standard CDM
(SCDM), a blue CDM model (BCDM) with primordial spectral index $n=1.2$ and a
blue Mixed Dark Matter model (BMDM), still with primordial spectral index
$n=1.2$, but with $\Omega_{CDM}=0.65$, $\Omega_{B}=0.05$ and
$\Omega_{\nu}=0.3$. We choose this last model, even if only marginally
consistent with the linear analysis of the velocity field presented in Section
3.1, in order to study an extreme case where the effects of the hot component
can be largely appreciated; moreover this choice allows a direct comparison
with the results of Klypin et al. (1993) who simulated a model with the same
$\Omega_\nu$ but with a primordial spectral index $n=1$.

We define the present time by fixing the normalization to the value implied by
{\it COBE} DMR data in the absence of any relevant gravitational wave
contribution: this corresponds to approximately $b=1$, $b=0.5$ and $b=0.9$, for
SCDM, BCDM and BMDM respectively.

In the case of CDM models (SCDM and BCDM), we used $128^3$ grid--points and
$128^3$ particles. Initial conditions were set using the standard approach
based on applying the Zel'dovich approximation and assuming Gaussian primordial
fluctuations with power--spectrum $P(k) \propto k^{n} T^2(k)$, where $T(k) = [1
+ 6.8 k + 72.0 k^{3/2} + 16.0 k^2 ]^{-1}$ is the CDM transfer function (Davis
\et 1985). More details about these simulations can be found in Moscardini et
al. (1995).

The presence of the neutrino component in BMDM was taken into account in the
initial conditions as follows. We used the transfer functions computed in
Section 3 for each component separately. At a redshift $z$ the neutrinos have a
thermal motion with velocity distribution given by the Fermi--Dirac statistics
\be
dn(v) \propto {v^2 dv \over {\exp[v/v_0(z)]+1}} ~,
\ee
where $v_0(z) = 7.2 (1+z) (m_\nu/7 {\rm eV})^{-1}$ km s$^{-1}$, $m_\nu$ being
the neutrino mass; the corresponding rms velocity is $v_{{\rm rms}}(z)=3.596
{}~v_0(z)$. We start our simulations at $z=20$.

Very rapidly the growth of baryonic fluctuations becomes similar to that of the
cold component: because of this reason (see also Klypin et al. 1993) we
preferred to include the baryons with the cold particles and considered only
two components. Our N--body simulation of BMDM includes three sets each having
$128^3$ particles, one for the cold particles and two for the hot ones; the
relative masses of cold and hot particles are 0.7 and 0.15 respectively. The
initial conditions were generated by following the same method used by Klypin
et al. (1993), still based on the Zel'dovich approximation. The displacement of
the cold particles was obtained with a power--spectrum where only the transfer
function of the cold plus baryonic components was considered. The same random
phases were used to generate the displacement of the hot particles: in this
case the power--spectrum contains only the transfer function for the hot
component. A thermal velocity with random direction and amplitude drawn from
the Fermi--Dirac distribution was added to the velocity of the hot particles.
Imposing that the two sets of hot particles have velocities with opposite
directions avoids the generation of spurious fluctuations.

\subsection{Matter distribution}

In Figure 4 we plot the projected particle positions from a slice of depth
$20~h^{-1}$ Mpc, for the three models at the present time. A 1/4 random
sampling is adopted and only the cold particles are shown in the case of BMDM.
At first glance, the resulting distributions for SCDM and BMDM are quite
similar, even if in the latter case larger underdense regions are present.
Structures are instead more evident for BCDM, where voids are larger, the mass
distribution being more lumpy with many isolated clusters.

An interesting insight on the clustering of the mass can be obtained from the
counts--in--cells analysis. In particular, we calculate the variance $\sigma^2
\equiv \langle \delta^2 \rangle $. Figure 5 shows $\sigma^2$ as a function of
the side $R$ of cubic cells for the three models at the present time;
shot--noise corrections were applied. For comparison, we also show the results
(with $95\%$ error bars) obtained by Loveday et al. (1992) in the analysis of
the Stromlo--APM galaxy redshift survey. According to Loveday et al. (1992)
galaxies in this catalog are nearly unbiased with respect to the mass; this
implies that their counts can be directly related to the matter distribution.
The results we show refer to real space: we checked that redshift distortions
in our simulations slightly affect the variance only on very small scales.

The SCDM model, as already known (Efstathiou et al. 1990; Loveday et al. 1992),
has too small a variance on scales ranging from 20 to 60 $h^{-1}$ Mpc. The
clustering for BCDM is too strong on small scales: $\sigma^2$ is outside the
$95\%$ observational range for $R \mincir 35 ~h^{-1}$ Mpc. On the contrary, the
results for BMDM are very close to the observational ones.

One of the most important features of the large--scale matter distribution is
the presence of large voids: for example, de Lapparent, Geller, \& Huchra
(1986) found in slices of the CfA survey empty regions of radius around 50
$h^{-1}$ Mpc. More recently, Vogeley et al. (1994a) compared void statistics of
the extended CfA survey with simulations of different cosmological models,
finding that models with Gaussian initial fluctuations and CDM--like primordial
power--spectra are favoured. Comparing the Perseus--Pisces survey with N--body
simulations, Ghigna et al. (1994) find that a MDM model with
$\Omega_{CDM}/\Omega_{\nu}/\Omega_{B} = 0.6/0.3/0.1$ and primordial spectral
index $n=1$, exceeds the observational Void Probability Function for radii
smaller than 10 $h^{-1}$ Mpc, while a CDM model fares better. Statistics aiming
at measuring the presence of voids in the universe are plagued by the very
ambiguity in the definition of `void'. For our analysis we preferred to
calculate the Underdensity Probability Function (UPF) $P_{80}(R)$, defined as
the probability of having a cubic region of side $R$ with a density more than
$80\%$ below the mean (i.e. $\delta<-0.8$). This statistic is less sensitive to
the estimate of the mean density and to shot--noise, and permits to probe
larger scales (being non--zero for larger $R$) than the usual Void Probability
Function. Figure 6 shows $P_{80}$ as a function of $R$. As expected, and in
agreement with the analysis of Piran et al. (1993), a blue primordial spectrum
increases the presence of large voids. In fact, while the probability for SCDM
goes to zero for $R \approx 25 ~h^{-1}$ Mpc, BMDM and BCDM have a non--zero
$P_{80}$ out to 30 and 40 $h^{-1}$ Mpc respectively. For small--scale voids,
SCDM and BMDM behave in a very similar way; on the contrary, as also evident
from the slice in Figure 4, BCDM presents also a large amount of small voids.

Another test of the matter distribution we consider is the mean {\it contour
genus} per unit volume $g_S$. Recently, Vogeley et al. (1994b), performing a
topological analysis of the CfA redshift survey, found that SCDM has too a
large amplitude on smoothing scales $\mincir 10 ~h^{-1}$ Mpc, while an open CDM
model and a CDM model with non--zero cosmological constant are always
consistent with the observed genus. Figure 7 shows, for the three models at the
present time, the $g_S$ curves obtained using the program by Weinberg (1988).
The simulated data have been filtered by a Gaussian window, $W(r)= (\pi^{3/2}
\lambda^3)^{-1} \exp(-r^2/\lambda_e^2)$, with two different radii,
$\lambda_e=6~h^{-1}$ and $\lambda_e=20~h^{-1}$ Mpc, in order to examine the
topology of structure on both nonlinear and linear scales. For the ease of
comparison with Vogeley et al. (1994b), we consider contours of fixed volume
fraction and we plot $g_S$ as a function of the threshold $\nu$ parameterizing
the volume fraction in a Gaussian field: $F_{vol}=(2\pi)^{-1/2} \int_\nu^\infty
\exp(-t^2/2) dt$. Moreover we normalize the genus curves using the volume of
the CfA1+2 survey. Error bars, obtained by a bootstrap technique, are always
smaller than 10\%; for clearness they are not shown in the figure. For a
Gaussian random field (e.g. Doroshkevich 1970),
\be
g_S=g_0 (1-\nu^2) \exp (-\nu^2/2)\,,
\ee
where the normalization $g_0$ is related to the power--spectrum of density
fluctuations $P(k)$ through
\be
g_0={1\over {4 \pi^2}} \left[
{\int dk k^4 P(k)  \over {3 \int d k k^2 P(k) }} \right]^{3/2} \,.
\ee
The action of gravity changes the underlying statistics for the density
fluctuation $\delta$, breaking the symmetry of $g_S$. A shift of the peak of
the genus curve towards either high or low density reflects a topology that is
bubble--like or meatball--like respectively. In order to quantify this shift,
we use a statistics $\Delta \nu$ (e.g. Park, Gott, \& da Costa 1992; Vogeley et
al. 1994b) defined as
\be
\Delta \nu = {{\int_{-1}^1 d \nu \nu g(\nu)_{obs} } \over
{\int_{-1}^1 d \nu g(\nu)_{fit} }}\,,
\ee
where $g(\nu)_{obs}$ and $g(\nu)_{fit}$ are the measured and the best--fit
random--phase genus curves respectively. In agreement with the previous
definition, $\Delta \nu$ is positive (negative) when the genus curve has a
bubble (meatball) shift. In Table 4 we compare the results for $g_0$ and
$\Delta \nu$ at the two filtering radii for our three models and the CfA1+2
sample. In agreement with Vogeley et al. (1994b), SCDM is found not to
reproduce the observed topology in the nonlinear regime: the amplitude of $g_0$
at $\lambda_e = 6 ~h^{-1}$ Mpc is too large and the bubble shift too small;
moreover, at $\lambda_e = 20 ~h^{-1}$ Mpc the peak shift is in the wrong
direction. A similar negative trend is shown by BCDM, both at small and large
smoothing. On the contrary, BMDM has the desired behaviour: the CfA1+2 data are
well inside the error bars.

\begin{center}
{\bf Table 4.} Topological statistics.
\end{center}
\begin{center}
\begin{tabular}{|l||c|c||c|c|}
\hline
\hline
 & \multicolumn{2}{c||}{ $\lambda_e=6 ~h^{-1}$ Mpc} &
 \multicolumn{2}{c|}{ $\lambda_e=20 h~^{-1}$ Mpc} \\
\hline
\hline
        & $g_0$          & $\Delta_\nu$  & $g_0$         & $\Delta_\nu$  \\
\hline
CfA1+2  & 7.70           & 0.19          & 4.00          & 0.14           \\
SCDM    & $11.37\pm2.09$ & $0.07\pm0.12$ & $3.98\pm1.30$ & $-0.13\pm0.14$ \\
BCDM    & $11.11\pm1.97$ & $0.05\pm0.13$ & $5.02\pm1.28$ & $-0.07\pm0.14$ \\
BMDM    & $ 8.47\pm1.75$ & $0.13\pm0.12$ & $3.77\pm1.09$ & $0.05\pm0.16$  \\
\hline
\end{tabular}
\end{center}

\subsection{Velocity field statistics}

In order to recover the velocity field from each simulation, we follow a
standard procedure (e.g. Kofman et al. 1994). First, we interpolate the mass
and momentum from the particle distribution onto a cubic grid with $128^3$
grid--points, using a Triangular Shaped Cloud algorithm (see e.g. Hockney \&
Eastwood 1981); we then further smooth by a Gaussian filter with width
$2~h^{-1}$ Mpc, to ensure a non--zero density at every grid--point. The
velocity at each grid--point is defined as the momentum divided by the mass
density.

The left panel of Figure 8 shows the velocity modulus $v \equiv |\vec v|$
distribution function $P(v)$ for our models. The blue models generally produce
larger flows. The difference is statistically significant: in BCDM and BMDM
there are respectively about 700 and 13,000 grid--points with a velocity
greater than 2000 km s$^{-1}$, whereas SCDM contains no grid--point with such a
high velocity.

To further characterize the velocity field of each model, we computed the bulk
flow, i.e. the amplitude of the center--of--mass velocity of a sphere:
$v_{bulk}(R) = \left|\sum_{i=1}^n \vec v_i\right|/n$, where the sum extends
over the $n$ grid--points falling within a distance $R$ from the center. In
order to compare our results with observational data from the POTENT analysis
(e.g. Dekel 1994) we smooth the simulated velocity field by a Gaussian filter
of radius 12 $h^{-1}$ Mpc. In particular, the bulk flow is calculated for each
model by randomly selecting 100 different grid--points from the simulations and
calculating the statistics in top--hat spheres, of radius ranging from 10 to
130 $h^{-1}$ Mpc, centered on them. The central panel of Figure 8 shows the
values obtained by averaging the results over the 100 estimates. Error bars,
which are in all cases less than $5\%$, have not been plotted for clearness.
Due to the large error bars on POTENT data, all three models are in good
agreement with observations. As expected, the larger flows in blue models
displayed by the $P(v)$ analysis imply higher bulk flows, in better agreement
with the POTENT central values. This is the reason of the higher likelihood
resulting for blue models in the linear theory analysis reported in Section
3.1. Of course, having larger data samples, such as the ``Mark III" compilation
(Willick et al. 1994), can help to increase the discriminatory power of this
statistical test on the large--scale velocity field.

Another relevant statistic for the peculiar velocity field is the {\em Cosmic
Mach Number}, $\cal M$ (Ostriker \& Suto 1990), defined as the ratio of the
bulk flow to the one--point velocity dispersion $\sigma_v$: ${\cal M}(R) =
v_{bulk}(R) / \sigma_v(R)$, where $\sigma_v^2(R) = \sum_{i=1}^n (\vec v_i -
\vec v_{bulk})^2/n$ and the sum is extended over the same grid--points. Since
the bulk flow is caused by density fluctuations on scales larger than the
sampled volume, while the velocity dispersion mostly depends on the power on
smaller scales, $\cal M$ actually measures the ratio of large to small--scale
power in the velocity field. The mean values of $\cal M$, obtained in the same
way as for the bulk flows, are also shown in Figure 8 (right panel), for the
three models at the present time (where error bars, always less than 10\%, are
not plotted). The different estimates are very close and always inside the
error bars. This result shows that the Cosmic Mach Number poorly discriminates
among different models: a similar conclusion was obtained in the study of
non--scale--invariant (Moscardini et al. 1995) and non--Gaussian (Lucchin et
al. 1995) CDM models.

\subsection{Cluster analysis}

In order to identify clusters in our simulations we adopted a standard method
(White et al. 1987; Jing et al. 1994; Jing \& Fang 1994; Croft \& Efstathiou
1994). As a first step, we adopt a friends--of--friends algorithm to extract
the groups of cold particles with a linking parameter equal to one fifth of the
mean separation of the cold particles. After finding the center of mass of the
groups, we calculate the total (i.e. cold plus hot) mass inside a sphere of
radius $r_{cl}=1.5 ~h^{-1}$ Mpc, roughly corresponding to the radius of Abell
clusters. When the distance between the centers of two cluster candidates is
smaller than $2 r_{cl}$, we eliminate from the list the cluster with the
smaller mass and we repeat the whole procedure using all particles not
belonging to `clusters', until the number of clusters does not change: in this
way we can take into account the possible presence of more clusters in
overdense regions and/or filaments (i.e. superclusters) which would be merged
by the percolation algorithm.

In Figure 9 we show the spatial distribution of clusters obtained in our
simulations. In this plot, we consider only the richest clusters, i.e. those
with a mass larger than a suitable threshold chosen to reproduce the mean
density of the ACO/Abell catalogue, corresponding to an intercluster separation
$d_c \approx 40 ~h^{-1}$ Mpc.

The cumulative mass function $n(>M)$, i.e. the number of clusters per unit
volume with a mass larger than $M$, is presented for the three models in Figure
10. The BCDM model has a larger number of rich clusters compared with SCDM and
BMDM, which behave in a similar way. A comparison with the observational
estimates can be done using the mass function derived by Bahcall \& Cen (1992)
using both optical (richness, velocities and the luminosity function of
galaxies in clusters) and X--ray data (temperature distribution of clusters).
All models present too a high mass function for larger masses. The
disagreement, already known in the case of SCDM (e.g. Bahcall \& Cen 1992; Jing
\& Fang 1994) is significant: for example, no object with mass larger than
$10^{15} ~{\rm M}_\odot$ is found in the Bahcall \& Cen (1992) data, while
these objects may be found in the simulations. However, the estimates of
Bahcall \& Cen (1992) should be reconsidered, because of some underlying
assumptions, such as the density profile or sphericity (e.g. Klypin \& Rhee
1994). Independent estimates of the cluster mass function, using masses derived
directly from the dynamics (i.e. virial evaluations), give lower values for the
mass function (Biviano et al. 1993). Data from Bahcall \& Cen (1992) and
Biviano et al. (1993) are also shown in Figure 10; in the latter case error
bars account for uncertainties both in the best--fit parameters and in the
cluster mean density. Considering the Biviano et al. (1993) data, the agreement
with BMDM and SCDM is definitely better, while the number of very rich cluster
is still a problem for BCDM model.

These results are in good agreement with the linear cluster analysis discussed
in Section 3.2. All the considered models tend to overestimate the cluster
abundance, as it is clear from the fact that the Press--Schechter formula fits
the Bahcall \& Cen (1992) data only provided the threshold is chosen to take
high values, such as $\delta_c \sim 2.5$, which is equivalent to an
artificially low normalization (high $b$) for these models. If, on the other
hand, the cluster abundance had been used as a constraint to fix the
power--spectrum normalization (as done by e.g. Pogosyan \& Starobinsky 1995), a
value of $\sigma_8$ in contrast with the amplitude required by {\it COBE} data
would have come out.

For our simulated cluster catalogs we compute also the two--point correlation
function $\xi_{cc}$. Observational estimates of $\xi_{cc}$ are well fitted by
the relation $\xi_{cc}= (r/r_0)^{-1.8}$. The amplitude of the correlation
function and its possible dependence on the cluster richness are rather
controversial: recent estimates of $r_0$ based on different samples range from
13 to 25 $h^{-1}$ Mpc (e.g. Nichol, Briel, \& Henry 1994 and references
therein). Figure 11 shows the results for the cluster catalogs in our
simulations. For clearness, error bars, obtained by bootstrap resampling, are
only shown for BMDM. As already known (see e.g. Bahcall \& Cen 1992), the
cluster correlation function for SCDM is too low. On the other hand, BCDM
cannot reproduce the right slope: in fact, this model is overclustered at small
scales and underclustered at larger scales. Only the results for BMDM are well
inside the observational range: in this case $r_0$ is approximately 16 $h^{-1}$
Mpc.

\section{Conclusions}

In this work we presented a detailed comparison of cold and mixed dark matter
models, with blue primordial power--spectra, with observational data on large
scales. In our analysis we used both linear and nonlinear techniques. Let us
stress that this new class of models, based on the assumption that $n>1$, has
the same level of naturalness as those based on any other choice of the
primordial spectral index, such as the usual scale--invariant $n=1$ value or
the tilted, $n<1$, ones. Being primordially scale--free our initial conditions
are more natural than those where the power--spectrum contains a built--in
characteristic scale. Blue types of spectra are suggested by the need of
reconciling the amount of power required on large (i.e. in the range $10 - 100$
Mpc) scales with the amplitude of perturbations implied by {\it COBE} data on
larger scales. An unavoidable consequence of this choice, in the frame of cold
dark matter models, is an excessive level of fluctuations on small (i.e.
$\mincir 1$ Mpc) scale. This is the strongest motivation for resorting to the
free--streaming effect of a hot dark matter component. Indeed, the models which
fare better in most of our analyses are the blue mixed dark matter (BMDM) ones.
Since our analysis assumes the {\it COBE} DMR data as a normalization
constraint, the first test to consider is the level of temperature fluctuations
on degree scale. De Bernardis, Balbi, \& Vittorio (1995) tested BMDM models
with a late, sudden reionization of the intergalactic medium against the ARGO
(de Bernardis et al. 1994), {\em COBE} DMR (Bennett et al. 1994),  MAX (Devlin
et al. 1994) and MSAM (Cheng et al. 1994) data set. Their likelihood analysis
indicates that mixed dark matter models with blue power spectra ($n \sim 1.2$)
and a reionization at $z \sim 30$ are most consistent with the data, even if
the standard model with $n=1$ and without late reionization is statistically
indistinguishable from the best model. It is not a problem to have  a late
reionization in the context of blue models. Indeed, they can have enough
small--scale power at early times to produce an early generation of structures,
leading to late reionization (e.g. Carr, Gilbert, \& Lidsey 1994). Related to
this property is also an anticipation of the galaxy and cluster formation
epochs compared to the scale--invariant MDM case (see, e.g., Lyth \& Liddle
1995).

The large--scale matter distribution emerging from our BMDM scenario has the
right properties to fit the existing observational data on various statistics,
such as the counts--in--cells analysis of Stromlo--APM galaxies and the
topology of the CfA redshift survey; thanks to the increased power on large
scales compared to SCDM and MDM, the large--scale voids in our simulations also
appear in good shape to reproduce the amount and size of those observed in
galaxy catalogs. We also computed, both within linear theory and by numerical
simulations, the large--scale peculiar velocity field, and performed a
statistical comparison with observational data from POTENT (e.g. Dekel 1994):
although the best model selected by a Likelihood analysis is the blue CDM one,
we find that BMDM models, with a small fraction of hot dark matter can
generally provide a reasonable agreement with the peculiar velocity data.

A problem which remains unsolved in the frame of all the models we considered
is related to the cluster abundance. The general tendency is to produce too
many massive clusters. Our analysis, in this respect, strongly differs from
that of Pogosyan \& Starobinsky (1995), who normalized their power--spectra to
fit the cluster abundance, as deduced by a Press--Schechter formula, with the
Bahcall \& Cen (1992) data. For us, instead, once the {\it COBE} DMR
normalization has been imposed on the models, we obtain the cluster mass
function as a {\it prediction} of our analysis. We believe, on the other hand,
that many aspects of the cluster distribution represent an open problem for all
{\it natural} (i.e. not containing {\it ad hoc} parameters) scenarios of
structure formation, whose solution would probably require a better modeling of
cluster formation in numerical simulations than it is presently possible by
collisionless codes.

\section* {Acknowledgments}
This work has been partially supported by Italian Ministero dell'Universit\`a e
della Ricerca Scientifica e Tecnologica and by Consiglio Nazionale delle
Ricerche (Progetto Finalizzato Sistemi Informatici e Calcolo Parallelo). The
staff and the management of the CINECA Computer Center (Bologna) are warmly
acknowledged for their assistance and for allowing the use of computational
facilities.

\newpage
\large
\begin{center}
\noindent {\bf References}
\end{center}
\normalsize

\begin{trivlist}
\item[] Antonuccio--Delogu, V., \& Colafrancesco, S. 1994, ApJ, 427, 72
\item[] Bahcall, N.A., \& Cen, R. 1992, ApJ, 398, L81
\item[] Bennett, C.L., et al. 1994, ApJ, 430, 423
\item[] Bertschinger, E., Dekel, A., Faber, S.M., Dressler, A., \&
Burstein, D. 1990, ApJ, 364, 370
\item[] Biviano, A., Girardi, M., Giuricin, G., Mardirossian, F., \&
Mezzetti, M. 1993, ApJ, 411, L11
\item[] Bond, J.R., Cole, S., Efstathiou, G., \& Kaiser, N. 1991, ApJ, 370, 440
\item[] Bonometto, S.A., Lucchin, F., Occhionero, F., \& Vittorio, N. 1984,
A\&A, 138, 477
\item[] Carr, B.J., Gilbert, J.H., \& Lidsey, J. 1994, Phys. Rev.,
D50, 4853
\item[] Carr, B.J., \& Lidsey, J. 1993, Phys. Rev., D48, 543
\item[] Cen, R., \& Ostriker, J.P. 1994, ApJ, 431, 451
\item[] Cheng, E.S., et al. 1994, ApJ, 422, L40
\item[] Colafrancesco, S., \& Vittorio, N. 1994, ApJ, 422, 443
\item[] Copeland, E. J., et al. 1994, Phys. Rev.,  D49, 6410
\item[] Croft, R.A.C., \& Efstathiou, G. 1994, MNRAS, 267, 390
\item[] Davis, M., Efstathiou, G., Frenk, C.S., \& White, S.D.M.
1985, ApJ, 292, 371
\item[] Davis, M., Summers, F.J., \& Schlegel, M. 1992, Nature, 359, 393
\item[] de Bernardis, P., et al. 1994, ApJ, 422, L33
\item[] de Bernardis, P., Balbi, A., \& Vittorio, N. 1995,
preprint
\item[] de Gasperis, G., Muciaccia, P.F., \& Vittorio, N. 1995, ApJ, 439, 1
\item[] Dekel, A. 1994, ARAA, 32, 371
\item[] de Lapparent, V., Geller, M.J., \& Huchra, J.P. 1986, ApJ, 302, L1
\item[] Del Grande, P., \& Vittorio, N. 1992, ApJ, 397, 26
\item[] Devlin, M., et al. 1994, preprint
\item[] Doroshkevich, A.G. 1970, Astrophysics, 6, 320
\item[] Efstathiou, G., Kaiser, N., Saunders, W., Lawrence, A.,
Rowan--Robinson, M., Ellis, R.S., \& Frenk, C.S. 1990, MNRAS, 247, 10P
\item[] Geller, M.J., \& Huchra, J.P. 1989, Science 246, 897
\item[] Ghigna, S., Borgani, S., Bonometto, S.A., Guzzo, L.,
Klypin, A., Primack, J.R., Giovanelli, R., \& Haynes, M.P. 1994, ApJ, 437, L71
\item[] Gioia, I.M., Maccacaro, T., Schild, R.E., Wolter, A., Stocke, J.T.,
Morris, S.L., \& Henry, J.P. 1990, ApJS, 72, 567
\item[] G\'orski, K.M., et al. 1994, ApJ, 430, L89
\item[] Gundersen, et al. 1993, ApJ, 413, L1
\item[] Hancock, S., Davies, R.D., Lasenby, A.N., Gutierrez de la Cruz,
C.M., Watson, R.A., Rebolo, R., \& Beckman, J.E. 1994, Nature, 367, 333
\item[] Hockney, R., \& Eastwood, J.W. 1981, Computer Simulation Using
Particles (New York: Mc Graw--Hill)
\item[] Hu, W., Scott, D., \&  Silk, J. 1994, ApJ, 430, L97
\item[] Jing, Y.P., \& Fang, L.Z. 1994, ApJ, 432, 438
\item[] Jing, Y.P., Mo, H.J., Borner, G., \& Fang, L.Z. 1994, A\&A, 243, 295
\item[] Klypin, A., Borgani, S., Holtzman, J., \& Primack, J. 1995, ApJ,
in press
\item[] Klypin, A., Holtzman, J., Primack, J., \& Reg\H{o}s, E. 1993, ApJ,
416, 1
\item[] Klypin, A., Nolthenius, R., \& Primack, J. 1995, preprint
\item[] Klypin, A., \& Rhee, G. 1994, ApJ, 428, 399
\item[] Kofman, L., Bertschinger, E., Gelb, J.M., Nusser, A., \&
Dekel, A. 1994, ApJ, 420, 44
\item[] Kogut, A., et al. 1993, ApJ, 419, 1
\item[] Kowalski, M.P., Ulmer, M.P., Cruddace, R.G.,  \& Wood, K.S.
1984, ApJS, 56, 403
\item[] Lauer, T., \& Postman, M. 1994, ApJ, 425, 418
\item[] Liddle, A.R., \& Lyth, D.H. 1993, MNRAS, 265, 379
\item[] Linde, A. 1991a, Phys. Lett., B249, 18
\item[] Linde, A. 1991b, Phys. Lett., B259, 38
\item[] Linde, A. 1994, Phys. Rev.,  D49, 748
\item[] Louis, W.C. 1994, in XVI Conf. on Neutrino Phys. \& Astrophys.,
Eilat, Israel
\item[] Loveday, J., Efstathiou, G., Peterson, B.A., \&
Maddox, S.J. 1992, ApJ, 400, L43
\item[] Lucchin, F., \& Matarrese, S. 1985, Phys. Rev., D32, 1316
\item[] Lucchin, F., Matarrese, S., Messina, A., Moscardini, L., \& Tormen,
G. 1995, MNRAS, 272, 859
\item[] Lyth, D.H., \& Liddle, A.R. 1995, Ap. Lett. \& Comm., in press
\item[] Ma, C.P., \& Bertschinger, E. 1994, ApJ, 434, L5
\item[] Mather, J. C., et al. 1994, ApJ, 420, 439
\item[] Mo, H.J., Jing, Y.P., \& Borner, G. 1993, MNRAS, 264, 825
\item[] Mollerach, S., Matarrese, S., \& Lucchin, F. 1994, Phys. Rev., D50,
4835
\item[] Moscardini, L., Tormen, G., Matarrese, S., \& Lucchin, F. 1995,
ApJ, in press
\item[] Nichol, R.C., Briel, U.G., \& Henry, J.P. 1994, MNRAS, 265, 867
\item[] Ostriker, J.P., \& Suto, Y. 1990, ApJ, 348, 378
\item[] Park, C., Gott, J.R., \& da Costa, L.N. 1992, ApJ, 392, L51
\item[] Peebles, P.J.E. 1981, ApJ, 248, 885
\item[] Piran, T., Lecar, M., Goldwirth, D.S., da Costa, L.N., \&
Blumenthal, G.R. 1993, MNRAS, 265, 681
\item[] Plionis, M., Borgani, S., Moscardini, L., \& Coles, P. 1995, ApJ,
441, L57
\item[] Pogosyan, D., \& Starobinsky, A. 1995, ApJ, in press
\item[] Press, W.H., \& Schechter, P.L. 1974, ApJ, 187, 425
\item[] Smoot, G.F., et al. 1992, ApJ, 396, L1
\item[] Vogeley, M.S., Geller, M.J., Park, C., \& Huchra, J.P. 1994a, AJ,
108, 745
\item[] Vogeley, M.S., Park, C., Geller, M.J.,
Huchra, J.P., \& Gott, J.R. 1994b, ApJ, 420, 525
\item[] Weinberg, D.H. 1988, PASP, 100, 1373
\item[] White, S.D.M., Efstathiou, G., \& Frenk, C.S. 1993, MNRAS, 262, 1023
\item[] White, S.D.M., Frenk, C.S., Davis, M., \& Efstathiou, G. 1987,
ApJ, 313, 505
\item[] Willick, J.A., Courteau, S., Faber, S.M., Burstein, D., \&
Dekel, A. 1994, preprint astro--ph/9411046
\item[] Zurek, W.H., Quinn, P.J., Salmon, J.K., \& Warren, M.S. 1994, ApJ,
431, 559
\end{trivlist}

\newpage

\section*{\center Figure captions}

\noindent
{\bf Figure 1.} The total transfer function (short--dashed line) and the
transfer function for baryonic (continuous line), cold (dotted line) and hot
components (long--dashed line), at different redshifts: a) $z=20$; b) $z=10$;
c) the present time $z=0$. The results refer to a flat MDM model with
$\Omega_\nu=0.3$, $\Omega_B=0.05$ and $h=0.5$.

\vspace{0.2truecm}
\noindent
{\bf Figure 2.} The rms value of the fluctuation density field at the present
epoch, $\sigma(R)$, is plotted versus the filtering scale $R$, in units of Mpc
($h=0.5$ has been used). The four curves refer to the different scenarios of
galaxy formation that we consider in this paper: BCDM (dotted curve), SCDM
(dot--dashed curve), BMDM (continuous curve) and MDM (dashed curve). All these
models have initial spectra normalized to the rms temperature fluctuation of
the {\it COBE} DMR maps.

\vspace{0.2truecm}
\noindent
{\bf Figure 3.} The predictions for the local X--ray luminosity function are
compared with data from Kowalski et al. (1984). The dotted curves in the three
panels are evaluated with the parameter choice listed in Table 3 which give the
best--fit of the data (see text for details). The dashed curves in each panel
are the XRLFs obtained with $\delta_c=1.7$ and fluctuation spectra normalized
to the COBE/DMR amplitude. X--ray luminosities are evaluated in the $2-6$ keV
energy band.

\vspace{0.2truecm}
\noindent
{\bf Figure 4.} Matter distribution in slices of 20 $h^{-1}$ Mpc for SCDM (top
panel), BMDM (left down) and BCDM (right down), at the present time. The box
refer to the whole simulation, 260 $h^{-1}$ Mpc. A random sampling of 1/4 is
adopted. For BMDM only the ``cold" particles are shown.

\vspace{0.2truecm}
\noindent
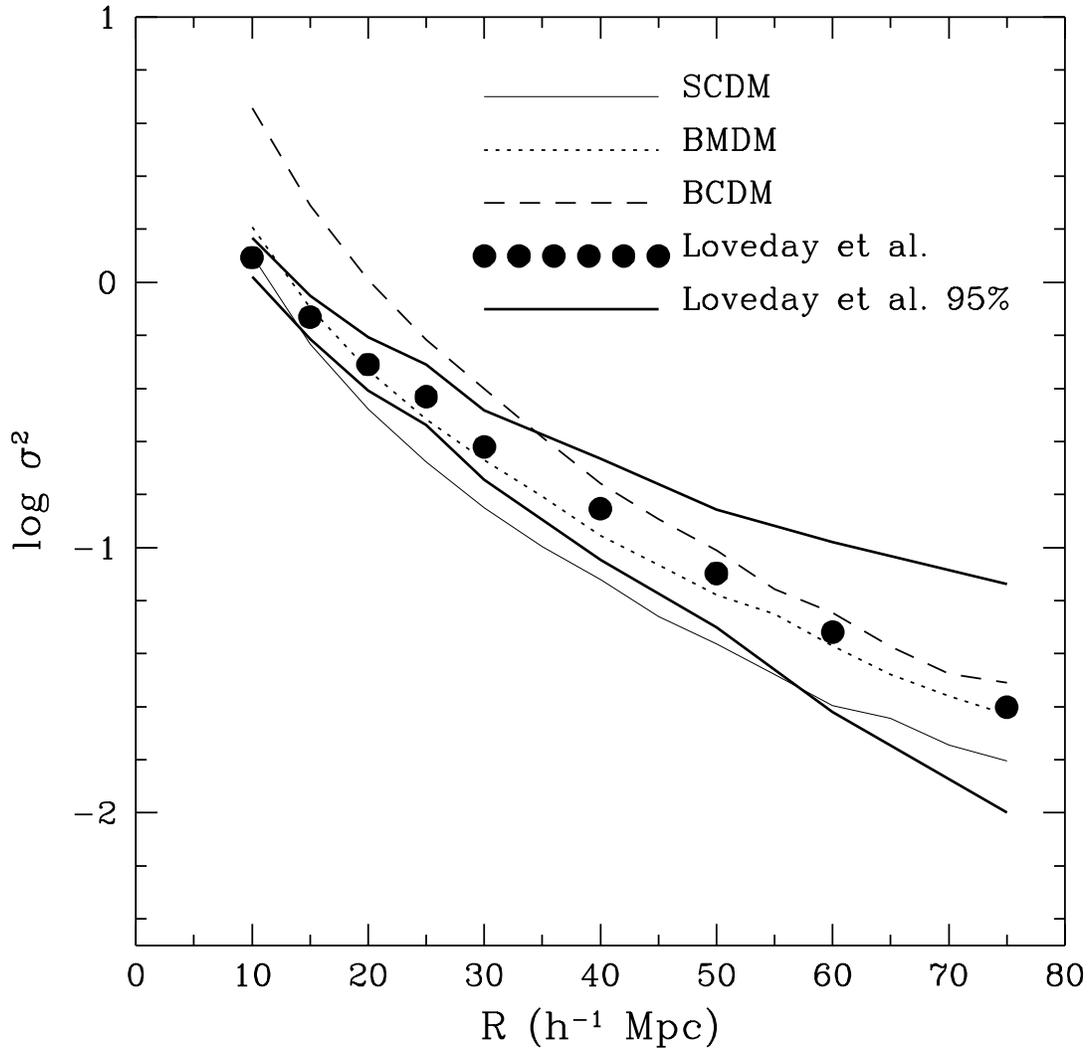
{\bf Figure 5.} Counts--in--cells analysis: the variance $\sigma^2$ as a
function of the side $R$ of cubic cells for SCDM (solid line), BMDM (dotted
line) and BCDM (dashed line). The filled circles and the heavy solid lines
refer to the Loveday et al. (1992) analysis of the APM--Stromlo catalog and
their 95\% confidence level, respectively.

\vspace{0.2truecm}
\noindent
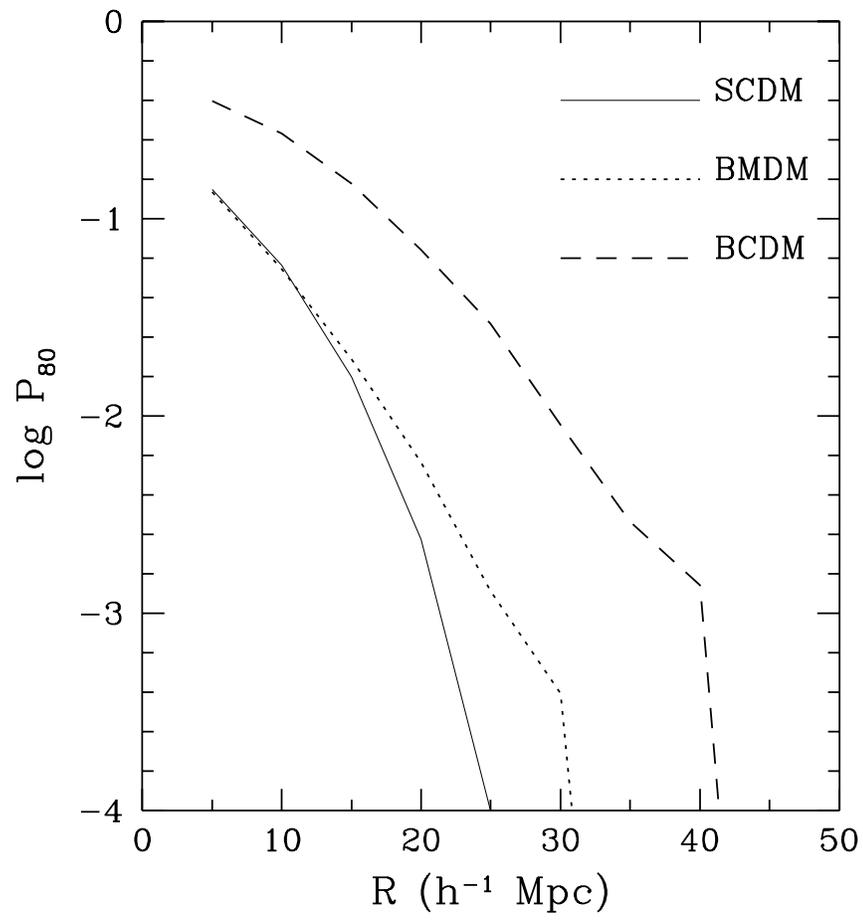
{\bf Figure 6.} The Underdensity Probablity Function $P_{80}$ as a function of
the side $R$ of cubic cells for SCDM (solid line), BMDM (dotted line) and BCDM
(dashed line).

\vspace{0.2truecm}
\noindent
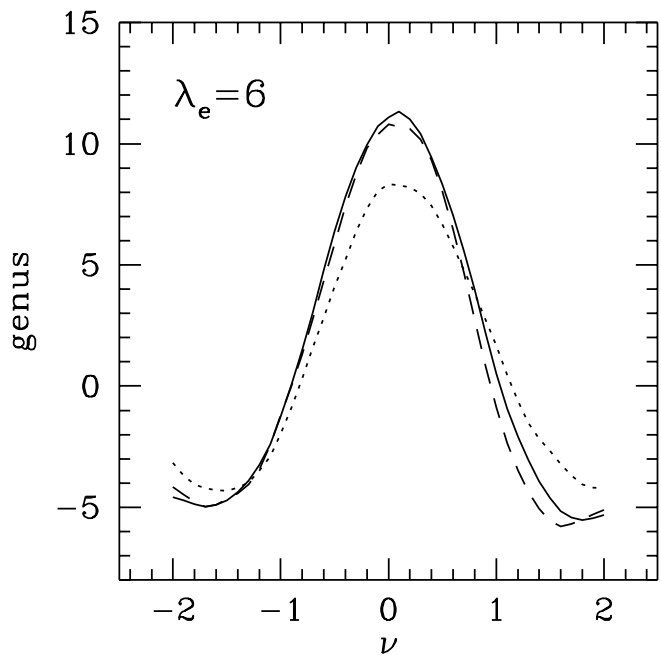
{\bf Figure 7.} The mean contour genus as a function of the effective threshold
$\nu$ for SCDM (solid line), BMDM (dotted line) and BCDM (dashed line). Two
different Gaussian filtering radii are considered: $\lambda_e=6 ~h^{-1}$ Mpc
(left panel) and $\lambda_e=20 ~h^{-1}$ Mpc (right panel).

\vspace{0.2truecm}
\noindent
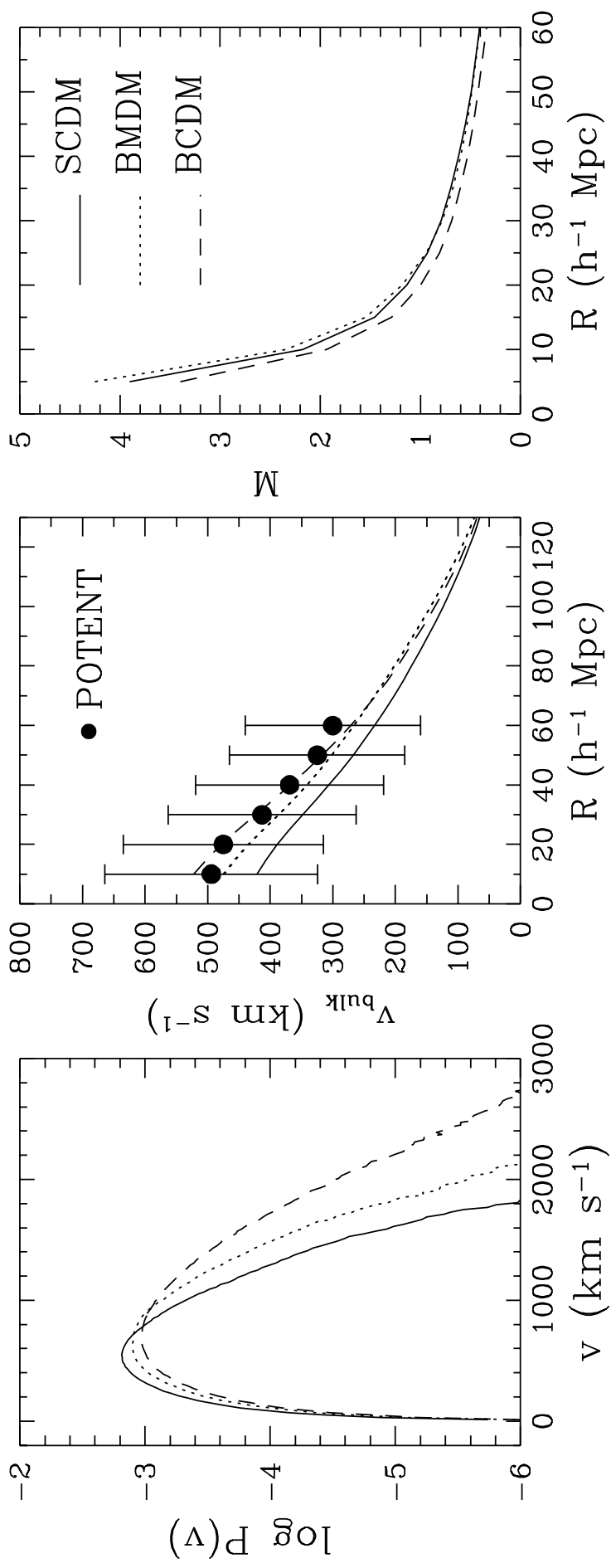
{\bf Figure 8.} Velocity field statistics. Different lines refer to different
models: SCDM (solid line), BMDM (dotted line) and BCDM (dashed line). Left
panel: the distribution function of the velocity modulus $v$. Central panel:
the bulk flow as a function of the radius $R$ of the sphere. Filled circles and
error bars refer to the POTENT data (Dekel 1994). Right panel: the Cosmic Mach
Number as a function of the radius $R$ of the sphere.

\vspace{0.2truecm}
\noindent
{\bf Figure 9.} Projected distribution of clusters identified with an
intercluster distance of $40 ~h^{-1}$ Mpc. The box represent the whole
simulation ($260 ~h^{-1}$ Mpc side): SCDM (left panel), BMDM (central panel)
and BCDM (right panel).

\vspace{0.2truecm}
\noindent
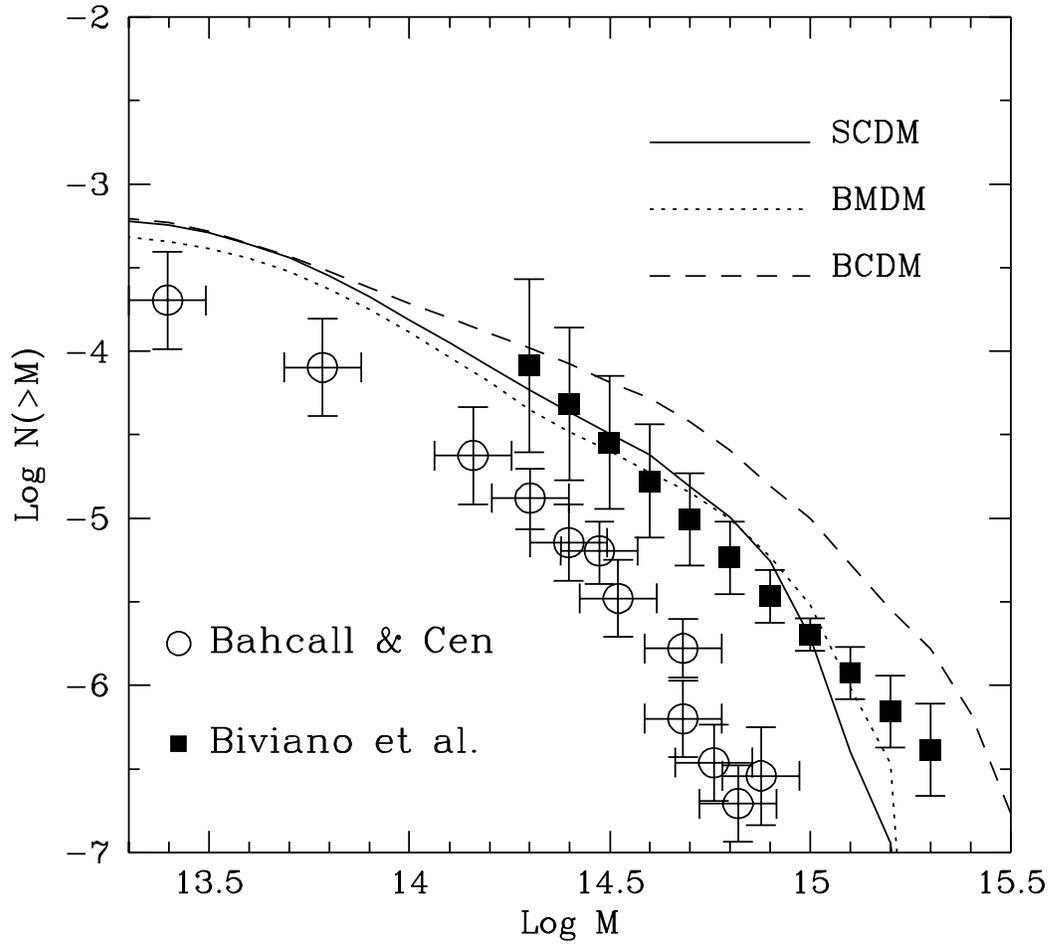
{\bf Figure 10.} The cluster mass function $N>(M)$ ($M$ in units of the solar
mass) for SCDM (solid line), BMDM (dotted line) and BCDM (dashed line). For
comparison the observational estimates by Bahcall \& Cen (1992) and Biviano et
al. (1993) are shown by open circles and filled squares, respectively.

\vspace{0.2truecm}
\noindent
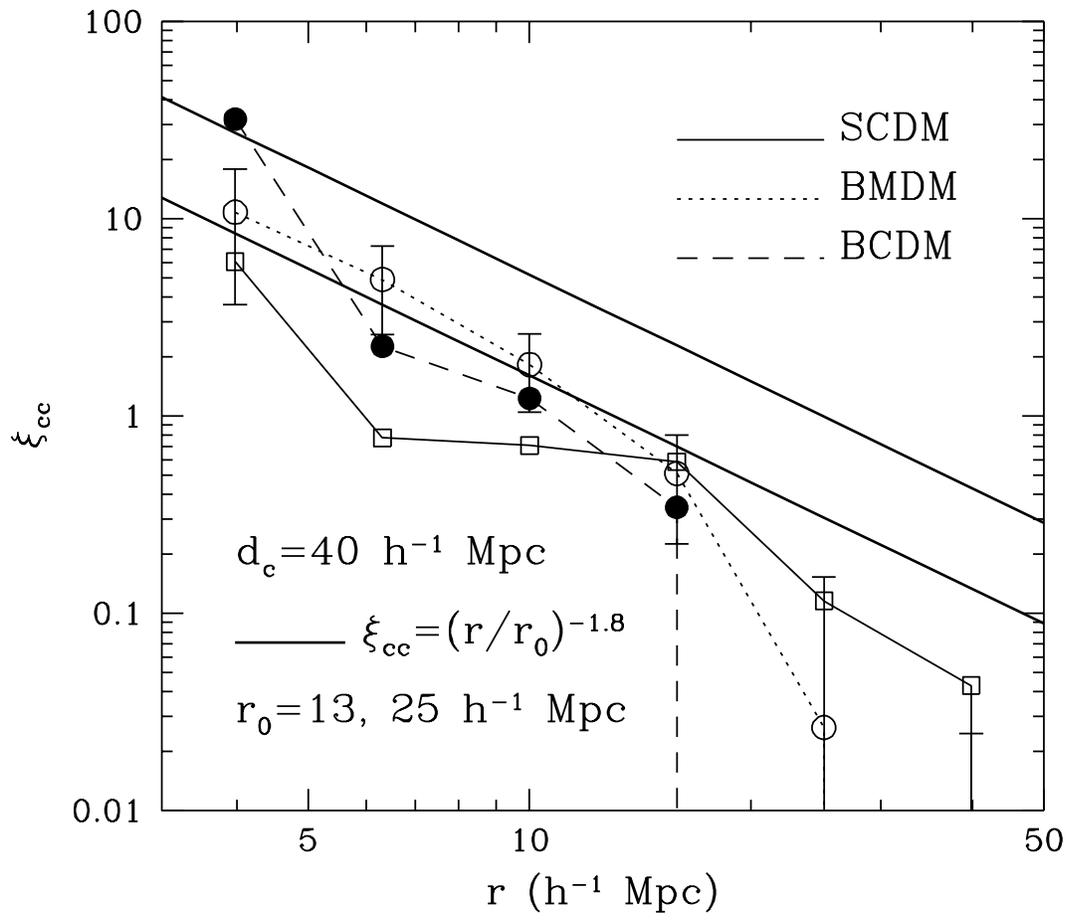
{\bf Figure 11.} The cluster correlation function $\xi_{cc}$ as a function of
the distance $r$ for SCDM (solid line and open squares), BMDM (dotted line and
open circles) and BCDM (dashed line and filled circles). Bootstrap errors are
shown only for BMDM. As comparison, the two heavy solid lines represent the
power--law $\xi_{cc}=(r/r_0)^{-1.8}$, with $r_0=13$ and $25 ~h^{-1}$ Mpc.

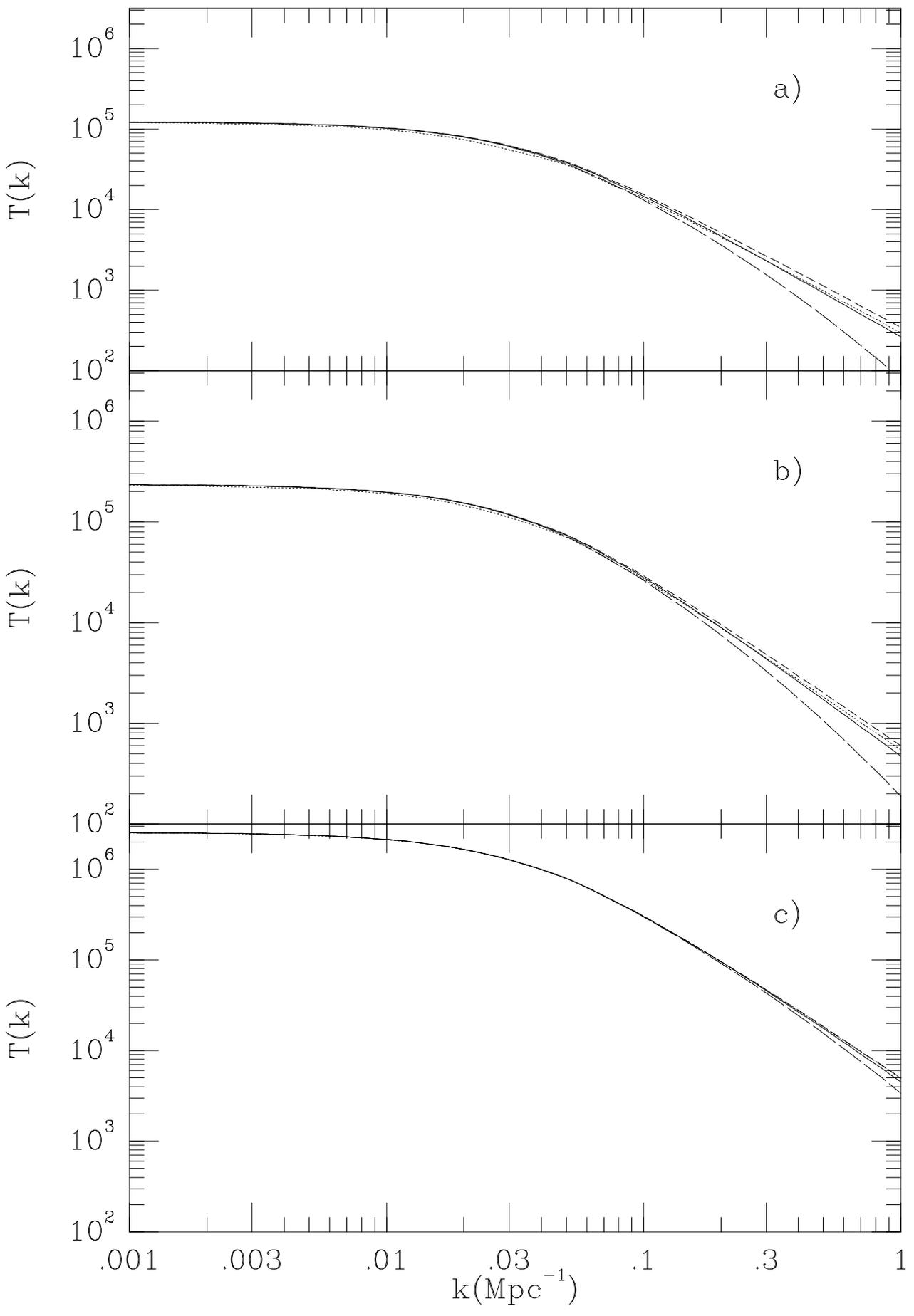
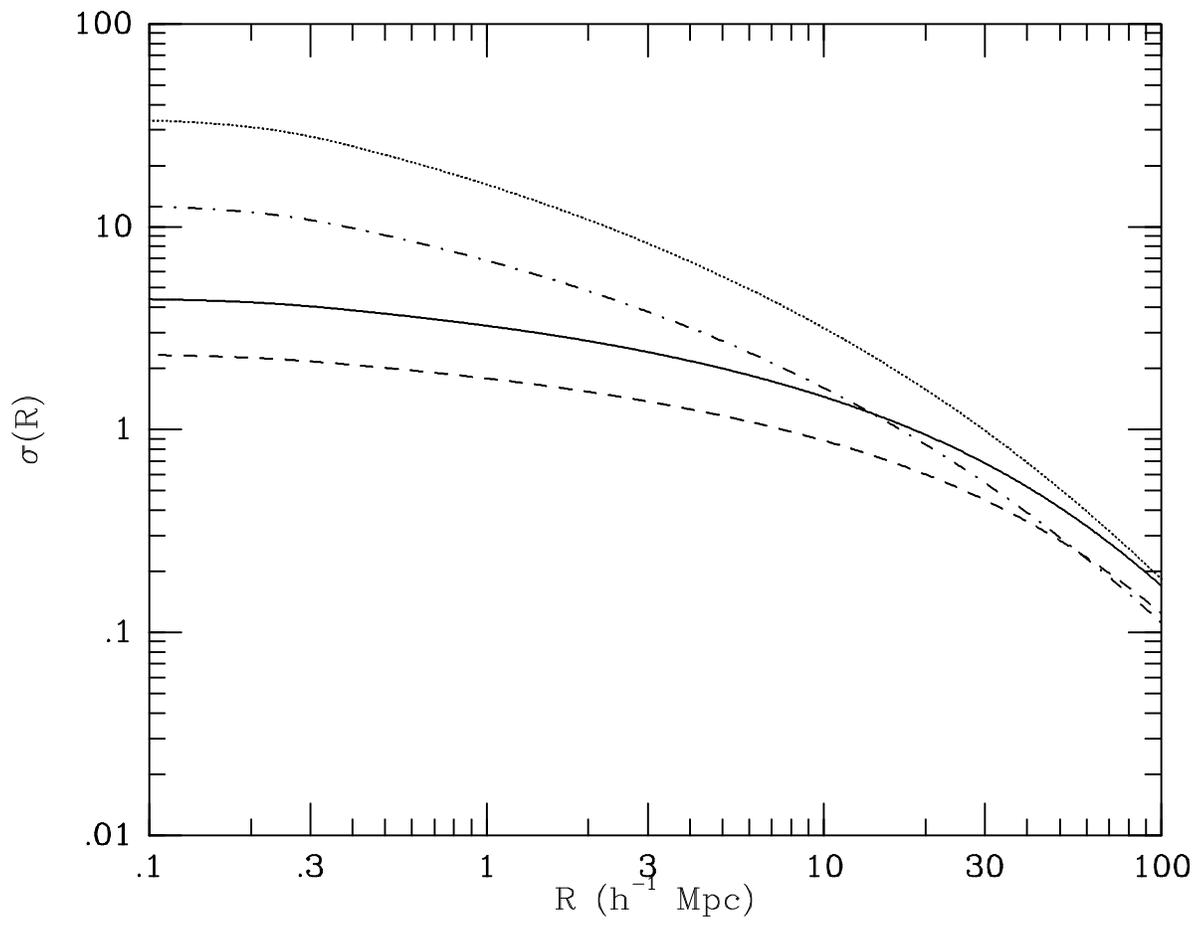
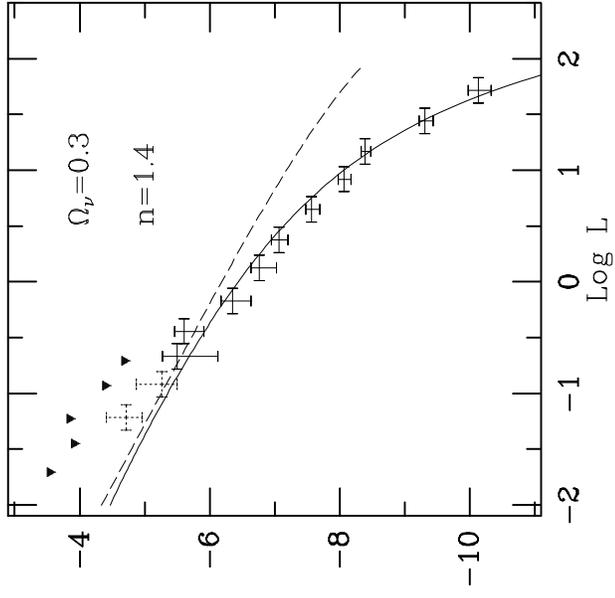
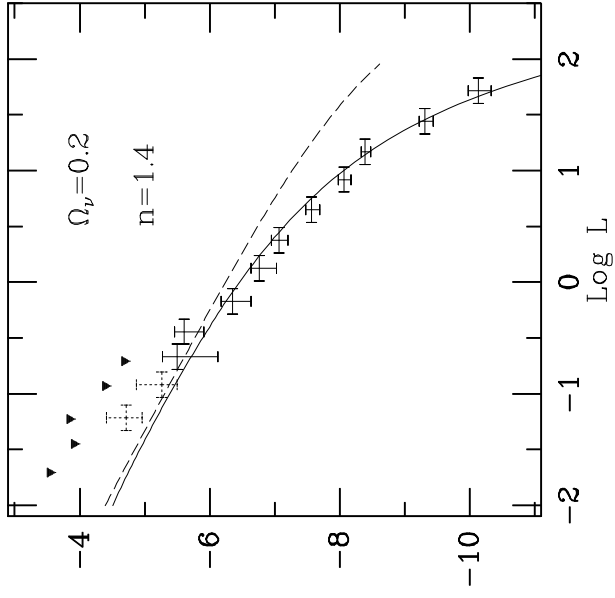
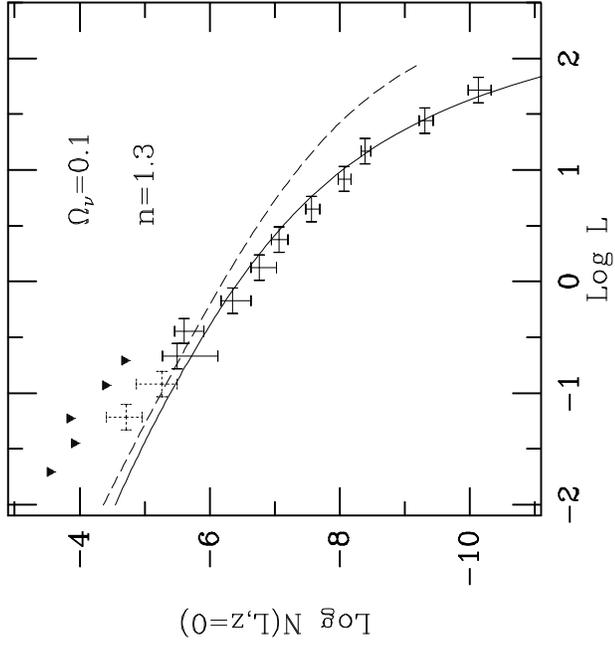
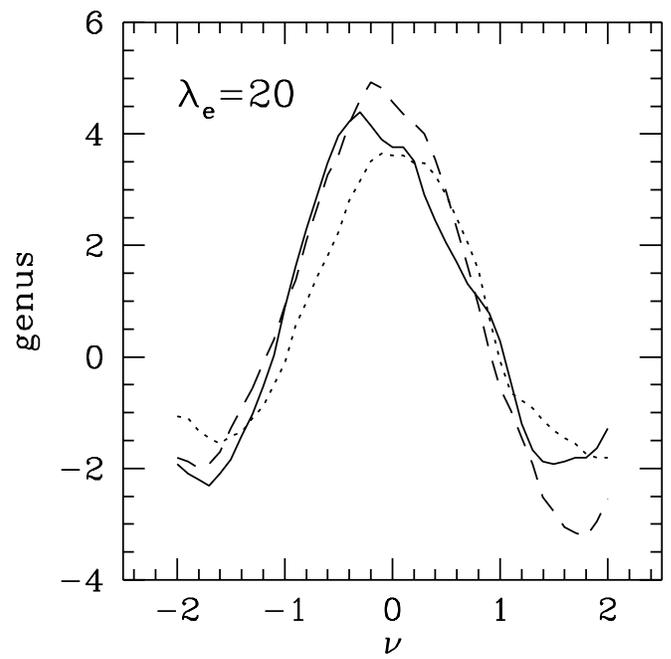
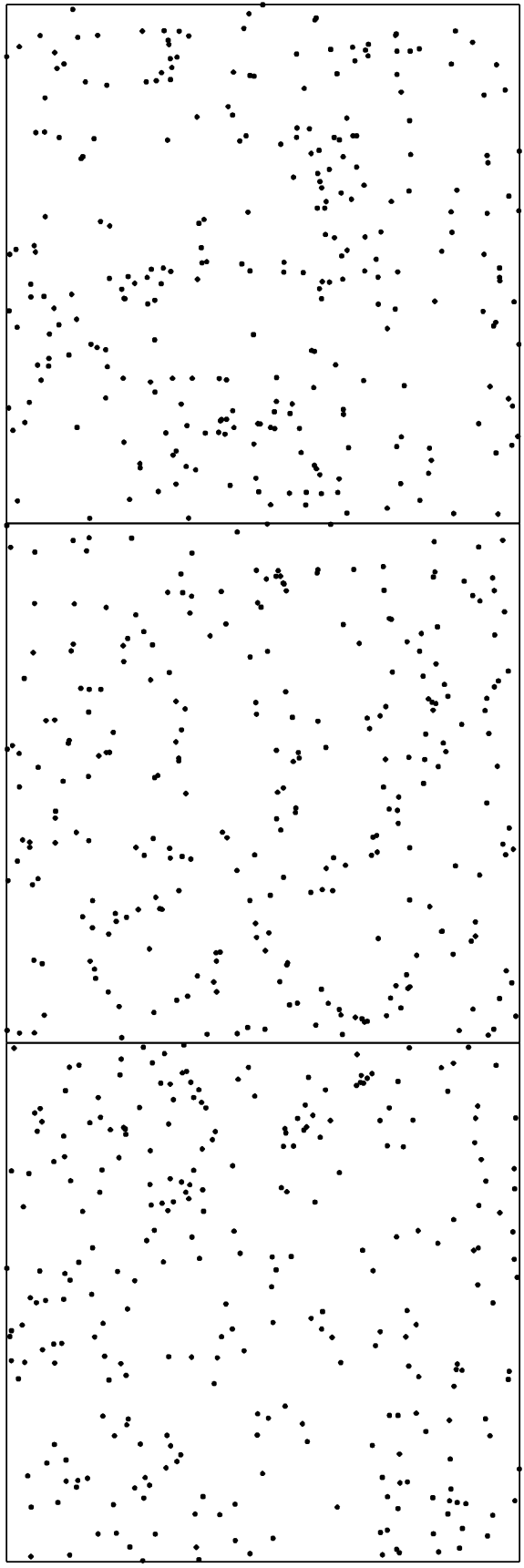
\end{document}